\documentclass[%
 reprint,
superscriptaddress,
 amsmath,amssymb,
 aps,
]{revtex4-2}

\usepackage{xcolor}
\usepackage{gensymb}
\usepackage{ulem}
\usepackage{xcolor}
\usepackage{graphicx}% Include figure files
\usepackage{dcolumn}% Align table columns on decimal point
\usepackage{bm}% bold math
\usepackage[hidelinks]{hyperref}
\usepackage{miller}
\newcommand{\cmmnt}[1]{\ignorespaces}
\begin{document}

\preprint{APS/123-QED}

\title{Electrical Manipulation of Telecom Color Centers in Silicon}

\affiliation{John A. Paulson School of Engineering and Applied Sciences, Harvard University, Cambridge, Massachusetts 02138, USA}
\affiliation{Department of Physics, Harvard University, Cambridge, Massachusetts 02138, USA}
\affiliation{AWS Center for Quantum Networking, Boston, Massachusetts 02135, USA}
\affiliation{Equal Contribution}

\author{Aaron M. Day}
    \affiliation{John A. Paulson School of Engineering and Applied Sciences, Harvard University, Cambridge, Massachusetts 02138, USA}
    \affiliation{Equal Contribution}
\author{Madison Sutula}
    \affiliation{Department of Physics, Harvard University, Cambridge, Massachusetts 02138, USA}
    \affiliation{Equal Contribution}
\author{Jonathan R. Dietz}  
    \affiliation{John A. Paulson School of Engineering and Applied Sciences, Harvard University, Cambridge, Massachusetts 02138, USA}
\author{Alexander Raun}  
    \affiliation{John A. Paulson School of Engineering and Applied Sciences, Harvard University, Cambridge, Massachusetts 02138, USA}
\author{Denis D. Sukachev}  
    \affiliation{AWS Center for Quantum Networking, Boston, Massachusetts 02135, USA}
\author{Mihir K. Bhaskar}  
    \affiliation{AWS Center for Quantum Networking, Boston, Massachusetts 02135, USA}
\author{Evelyn L. Hu}  
    \affiliation{John A. Paulson School of Engineering and Applied Sciences, Harvard University, Cambridge, Massachusetts 02138, USA}

\date{\today}
\begin{abstract}
Silicon color centers have recently emerged as promising candidates for commercial quantum technology, yet their interaction with electric fields has yet to be investigated. In this paper, we demonstrate electrical manipulation of telecom silicon color centers by fabricating lateral electrical diodes with an integrated G center ensemble in a commercial silicon on insulator wafer. The ensemble optical response is characterized under application of a reverse-biased DC electric field, observing both 100\% modulation of fluorescence signal, and wavelength redshift of approximately 1.4~GHz/V above a threshold voltage. Finally, we use G center fluorescence to directly image the electric field distribution within the devices, obtaining insight into the spatial and voltage-dependent variation of the junction depletion region and the associated mediating effects on the ensemble. Strong correlation between emitter-field coupling and generated photocurrent is observed. Our demonstration enables electrical control and stabilization of semiconductor quantum emitters.
\end{abstract}
\maketitle

\section{Introduction}

Silicon is a foundational material enabling applications across computation, electronics, and photonics. It is therefore intriguing to consider it as a host for quantum information processing applications. Although color centers in solids have emerged as a promising quantum memory platform, the most mature color center technologies \cite{bhaskar2020, knall2022, Stas_2022, Pompili_2021, rosenthal2023, miao2020, anderson2019, lukin2020sicoi, Babin2022, Day2023} are hosted in materials that are difficult to fabricate, such as diamond and silicon carbide. Recently, progress studying the G and T centers has renewed interest in using silicon color centers as quantum emitters \cite{Beaufils2018, bergeron2020,durand2021,higginbottom2022}. Additionally, demonstrations of silicon color center nanophotonic integration \cite{komza2022,redjem2023,prabhu2023,deabreu2023,lee2023} reveal the potential to leverage the long history of scalable device engineering in silicon to realize useful quantum technologies. Yet, better understanding of material processing is needed to achieve high yield and reproducible  formation of single G and T centers. 

Up to now, efforts to integrate color centers with silicon devices for quantum information applications have focused on nanophotonics. However, integrating quantum memories with electronic devices offers potential benefits in linewidth-narrowing \cite{anderson2019}, Stark tuning \cite{anderson2019,Rühl2019, Lukin2020}, charge state control \cite{de2017, widmann2019}, and readout \cite{niethammer2019}. 

\begin{figure}[h!]
\includegraphics[scale = 1]{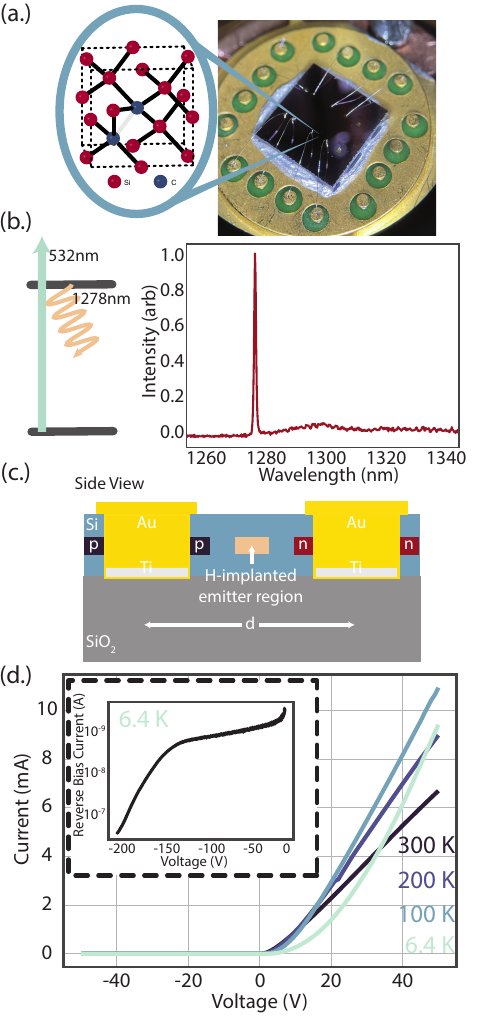}
\caption{\textbf{Diode-integrated silicon G centers} \textbf{(a.)} Carbon-related silicon color centers are integrated into lateral p$^+$-p-n$^+$ junctions (diodes) fabricated in silicon on insulator and electrically driven by a wire-bonded 16-pin helium cryostat connector. \textbf{(b.)} The color centers are optically excited by a 532~nm laser and fluoresce at 1278~nm in the telecommunication O-Band. \textbf{(c.)} Side profile of fabricated diodes. P- and n-doping is achieved via ion implantation, and hydrogen is locally incorporated to selectively form G centers at the junction center. \textbf{(d.)} Current-voltage (IV) curves of packaged diodes with integrated G center ensemble, measured as cryostat cools to base operating temperature of approximately 6~K. Inset shows low reverse bias leakage current, passing $-0.5~\mu A$ at -200~V.}
\label{Intro}
\end{figure}

In this article, we investigate the cryogenic optical response of a silicon color center to an applied electric field by integrating an ensemble of G centers with lateral p$^+$-p-n$^+$ diodes fabricated in silicon on insulator (SOI) (Fig. \ref{Intro}a). The G center--comprised of two substitutional carbon atoms bonded to an interstitial silicon atom--is an optically-active O-band emitter (Fig. \ref{Intro}b). Hydrogen implantation was found to be necessary for formation of G centers within our devices, and an ensemble is thereby localized to the middle of the diode junction by implanting hydrogen ions with a lithography-defined mask (Fig. \ref{Intro}c) in a wafer previously blanket-implanted with carbon. Above a spatially-dependent threshold voltage, the ensemble zero phonon line (ZPL) experiences a redshift up to 100~GHz at a rate of approximately 1.4~GHz/V. Additionally, we observed the continuous reduction of the G center optical fluorescence with increasing reversed bias voltage, and at -210V the fluorescence was fully suppressed. Finally, we employ the observed emitter-field coupling to image the spatial distribution of the electric field within the junction. 

The resultant spatial dependence of ZPL tuning and ensemble extinction suggest these mechanisms could be attributed to a combination of the Stark effect and Fermi-level shifting via band-bending. Our method has broad applicability for future control in quantum networking experiments, and serves as a tool for probing fundamental color center behavior. This approach is readily extensible to probe and control other color centers in silicon, and color centers in a wide range of semiconductor platforms which are easily doped, such as silicon carbide. 

\section{Lateral diodes with integrated G center ensemble}
We sought to realize a spatially-isolated G center ensemble maximally interacting with an electrical diode at a buried plane which is ultimately compatible with integrated silicon photonics. The maximum optical mode concentration of photonic crystal cavities in a 220~nm silicon layer would reside at 110~nm, thus we implement a design and fabrication strategy to support future hybrid electrical-optical coupling of semiconductor quantum emitters. To facilitate this, an industry standard 220~nm SOI wafer was utilized, with a dopant-defined diode embedded at a depth of 110~nm. Ion implantation combined with successive aligned optical lithography writes enabled masked localized incorporation of the p- and n- dopants, and the G center ensemble, at the desired depth. An etch-defined metallization strategy was employed to ensure robust electrical contact and performance at the dopant plane, and the device was packaged for cryogenic characterization. The current-voltage characteristics of the devices (Fig. \ref{Intro}d) do not degrade with temperature or masked hydrogen implantation, and they exhibit low leakage current under high reverse bias (Fig. \ref{Intro}d inset). 

\subsection{Device Design and Fabrication}
\begin{figure*}[ht!]
\includegraphics[scale = 1]{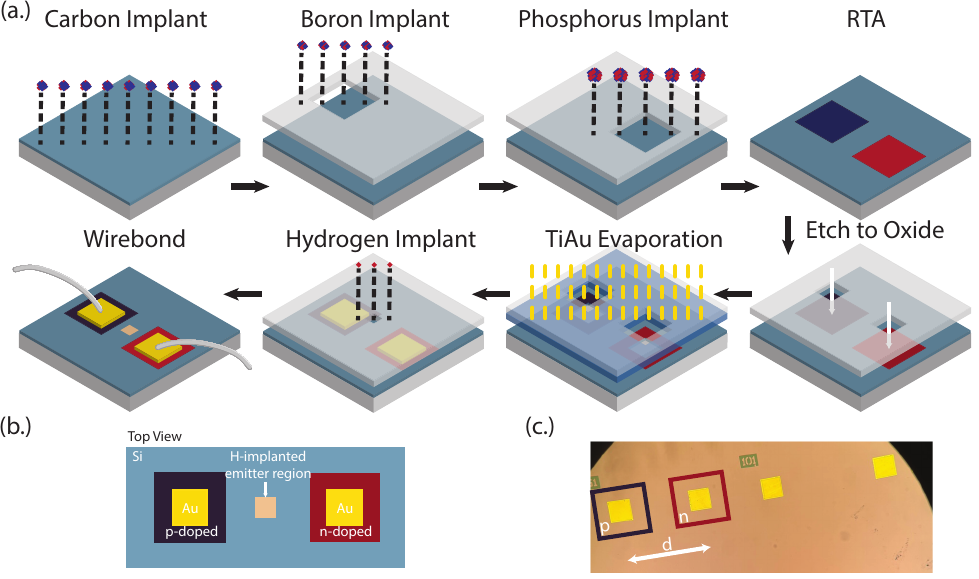}
\caption{\textbf{Device design and fabrication} \textbf{(a.)} Fabrication process for realizing diode-integrated G centers in SOI. \textbf{(b.)} Top view illustration of completed electrical device, and \textbf{(c.)} associated optical micrograph depicting appearance of finished device, with notable regions denoted. The junction spacing \textit{d} is varied across the wafer to enable a range of emitter-field coupling strengths. }
\label{Fab}
\end{figure*}

Lateral diodes are fabricated in commercial SOI (University Wafer, 220~nm Si on 2~$\mu$m buried oxide insulator, boron-doped, $\rho=10-20~\Omega$~cm, $\hkl<100>$ orientation) to facilitate simultaneous cryogenic optical and electrical measurement of color centers. The starting substrate of the devices is lightly p-doped based on prior reports of emitter synthesis \cite{prabhu2023, higginbottom2022}, though electrical performance would be improved in intrinsic material. The device design enables ease of optical access, variable junction width, and wafer scale--where hundreds of devices with swept parameters can be fabricated on a single commercial wafer defined via optical lithography. Further, the design co-locates the formed color centers and dopant-defined junction in the same spatial plane, improving emitter-field interaction. Device performance is validated with the COMSOL Multiphysics Semiconductor Module (see SI). The full device design and fabrication is depicted in Fig. \ref{Fab}a. An top-down diagram of the relevant regions of the device is illustrated in Fig. \ref{Fab}b, accompanied by an optical image of the finished devices (Fig. \ref{Fab}c). The full details of device fabrication are given in Methods. 

\subsection{Ensemble Incorporation}
To form G centers within the diodes, we first implant Si with carbon ions then rapid thermal anneal at 1000$^{\degree}$C to heal lattice damage. However, in contrast to some previous work \cite{prabhu2023,komza2022,hollenbach2022,redjem2023}, we did not observe G centers at this stage. Consistent with the findings of other works arguing proton irradiation facilitates incorporation of interstitial carbon into G centers \cite{baron2022,Beaufils2018}, we investigated varied means of hydrogenation to complete the G center formation (see SI \cite{SI}). Masked ion implantation of hydrogen was ultimately selected for the device-emitter integration to obtain a bright localized ensemble at the targeted depth where the electric field is strongest, with negligible degradation of electrical performance. Consistent with the findings of hydrogen's role in G center formation and stabilization, we found ensemble emission localized only to the implantation mask. Additionally, we found the G center to be unstable above 200$^{\degree}$C \cite{SI}, consistent with previous work \cite{Canham1987}, therefore requiring hydrogen incorporation to be the final fabrication step. Thus to ensure the diode fabrication was compatible with G center production, steps for both fabrication processes were interspersed. 

\section{Electrical manipulation}
Applying reverse-bias to a diode yields several electrical signatures: Fermi level engineering via band bending, generation of a depletion region with a local electric field internal to the junction, and suppression of leakage current. For junction-integrated color centers, band bending and junction depletion modulate the observed photoluminescence spectra, while low leakage current ensures minimal local heating. We first characterized the optical response of both the silicon free-exciton and the diode-integrated G center ensemble localized to the hydrogen implantation aperture under application of a reverse-bias DC electric field. Then the distribution of ensemble optical response is investigated across the junction, where competing effects from band bending and junction depletion can be distinguished. 

\subsection{Reverse bias}
\begin{figure*}[ht!]
\includegraphics[scale = 1]{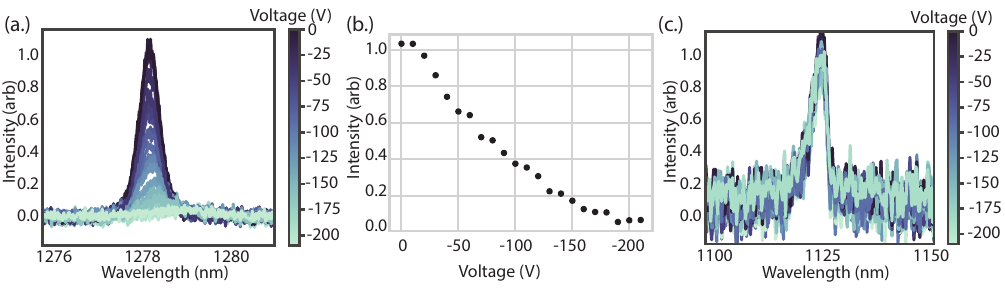}
\caption{\textbf{Reverse-bias electrical manipulation}  \textbf{(a.)} Optical response of G center ZPL under application of a DC electric field reverse biased from 0 to -210~V. Intensity continually decreases with increased reverse bias while the center wavelength shifts approximately 100~GHz to the red. \textbf{(b.)} Analysis of ZPL modulation ratio as a function of reverse bias, yielding 100\% modulation at -210~V. Normalized to ZPL intensity at zero bias. \textbf{(c.)} Preservation of silicon free-exciton under reverse bias. Unlike the G center, the exciton intensity, center wavelength, and linewidth are not correlated to increased reverse bias, indicating thermal effects are not a significant source of the observed G center behavior.}
\label{RB}
\end{figure*}
The G center ensemble response to a reverse-biased DC electric field is shown in Fig. \ref{RB}a-b. Photoluminescence was measured while sweeping the reverse bias in 10~V intervals from 0 to -210~V (Fig. \ref{RB}a). The fluorescence intensity of the ensemble reduced as a function of increasing reverse bias until the signal dropped below the noise floor of the measurement (Fig. \ref{RB}b). A 100~GHz redshift at a rate of approximately 1.4~GHz/V was observed in the G center ZPL above a spatially-dependent threshold voltage (explored in detail in Section III B). Additionally, the ensemble linewidth broadened as the center wavelength redshifted (see SI \cite{SI}). Both the ensemble redshift and linewidth broadening rates exhibit discontinuities from an ideal linear trend--this could be explained by the distribution of Stark shifts for each emitter within the confocal spot due to the varied dipole alignments within the ensemble. Future experiments with single emitters may elucidate whether the observed linewidth broadening and shifting is dominated by the local electric field interaction or via band-bending.

Reverse biased current of -0.5~{\textmu}A was passed at -200~V--corresponding to an applied power of 100~{\textmu}W spread over a 103~{\textmu}m junction gap. The device maintained low leakage current at high reverse bias, thus local heating is unlikely to be the source of the observed G center broadening, shifting, and modulation. To illustrate this point, the evolution of the silicon free-exciton line was investigated under the same bias conditions (Fig. \ref{RB}c). The silicon free-exciton is suppressed at elevated temperature (Supplementary Information \cite{SI}) and thus served as a probe of local junction heating. The exciton photoluminescence (PL) measured from 0 to -210~V is shown in Fig. \ref{RB}c. The exciton luminescence was not modified under reverse bias, consistent with the absence of significant heating--with fluctuations attributable to noise in the experiment. These results of the G center and silicon free-exciton are contrasted with the behavior under application of a high power forward bias in the supplementary information \cite{SI}. 

\subsection{Spatial Distribution of Emitter-Field Coupling}
Capturing the distribution of emitter optical response across the junction can aide in characterizing the nature of the emitter-field interaction. Band bending is achieved simply by making the p-contact increasingly negative and should be evident across the junction to varying degree. The Fermi level is a critical factor in a color center's optical activity \cite{son2021,zhang2023,pederson2023}, and thus an ensemble brightness gradient is expected to be present across the junction in proportion to the band bending achieved at a given voltage if the energy required for ionization is commensurate with the supplied electric potential \cite{Song1990,Udvarhelyi2021}. Conversely, excess carriers in the junction prevent complete depletion below a critical threshold voltage \cite{anderson2019,Candido2021}. The starting substrate of these devices is lightly p-doped and thus the depletion region is expected to be nonuniform, reaching the ensemble at a sufficiently large threshold voltage emerging first near the n-contact. Stark effect is mediated by a local electric field experienced by the ensemble, and thus is observed only when the depletion region reaches the ensemble \cite{anderson2019,Candido2021}. 

\begin{figure*}[t!]
\includegraphics[scale = 1]{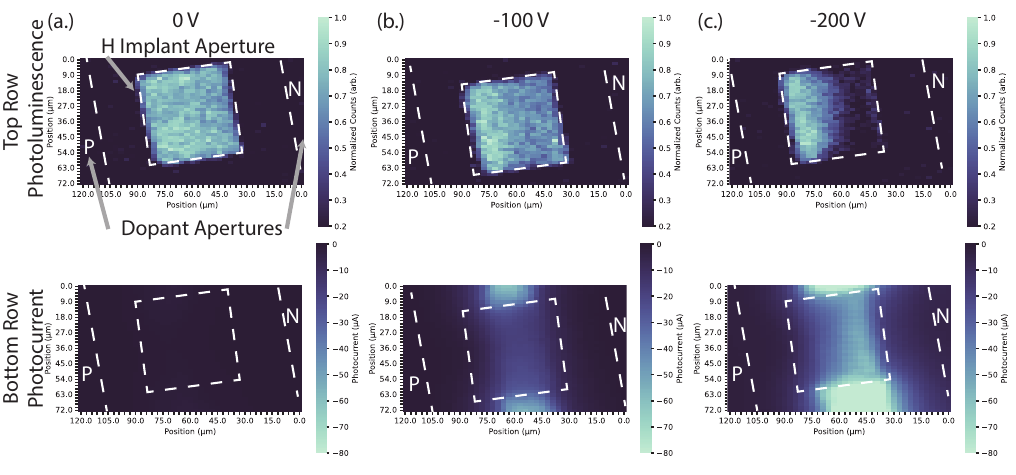}
\caption{\textbf{Spatial distribution of emitter-field interaction} Confocal maps of diode under reverse bias, depicting ensemble photoluminescence (top row) and measured junction photocurrent (bottom row) at \textbf{(a.)} 0~V, \textbf{(b.)} -100~V, and \textbf{(c.)} -200~V. The junction depletion region distribution is illustrated via the ZPL optical intensity modulation spreading across the ensemble beginning from the region closest to the n-type contact, with strong agreement found by correlating the optical response with injected photocurrent measured in the junction. PL color bar noise floor is set to 0.2 due to 1~s integration. Modulation ratio is obtained from normalizing to mean ZPL intensity at zero bias from (a.) top.}
\label{Confocal}
\end{figure*}

Additionally, the depletion region of the junction can be monitored directly by measuring confocal photocurrent, and hence one can correlate the depletion region's spatial occurrence confirmed via photocurrent with an optical response of the ensemble. To this end, the spatial distribution of the electric-field coupling to the G center ensemble is imaged (Fig. \ref{Confocal} top row), and correlated with the associated optically-generated photocurrent (Fig. \ref{Confocal} bottom row) of the diode under 0-bias (Fig. \ref{Confocal}a), -100~V (Fig. \ref{Confocal}b), and -200~V (Fig. \ref{Confocal}c.).

At 0~V, the localization of the G center ensemble is clear (Fig. \ref{Confocal}a top). G center PL is only observed in the 50$\times$50$~\mu$m aperture at the center of the diode through which hydrogen was implanted. As expected under zero bias, the measured photocurrent is negligible (Fig. \ref{Confocal}a bottom).

The confocal scan was repeated across the junction at a reverse bias of -100~V (Fig. \ref{Confocal}b). At -100~V, the optical intensity modulation ratio of the G centers was spatially dependant, with the emitters in the portion of the hydrogen-implant aperture closest to the n-contact showing 40\% greater suppression in response to the applied electric field than those nearest the p-contact. Interestingly, when comparing the confocal PL (top) with the associated confocal photocurrent measured in the device (bottom), the presence of the ensemble--and thus hydrogen--decreased the current passage across the junction, as the regions within the junction above and below the implant aperture demonstrated higher photo-responsivity.

Finally, the confocal PL and photocurrent spatial scan was repeated at a reverse bias of -200~V (Fig. \ref{Confocal}c). Closer to the n-type contact, 100\% modulation of the G center fluorescence is observed (Fig. \ref{Confocal}c top). Conversely, closer to the p-contact, G centers are minimally suppressed. Furthermore, the associated confocal photocurrent (Fig. \ref{Confocal}c bottom) follows the same spatial pattern. Optically-generated photocurrent measured in the junction was maximum in the region where emission is maximally modulated, confirming that the region of greatest depletion corresponds with strongest emitter interaction. As the strength of the reverse bias field increased, the spatial extent of the ensembles experiencing greatest optical modulation spread from the n-contact toward the p-contact as electrons and holes are swept toward their respective n- and p- contacts (Fig. \ref{Confocal} a-c). 

\begin{figure}[h!]
\includegraphics[scale = 1]{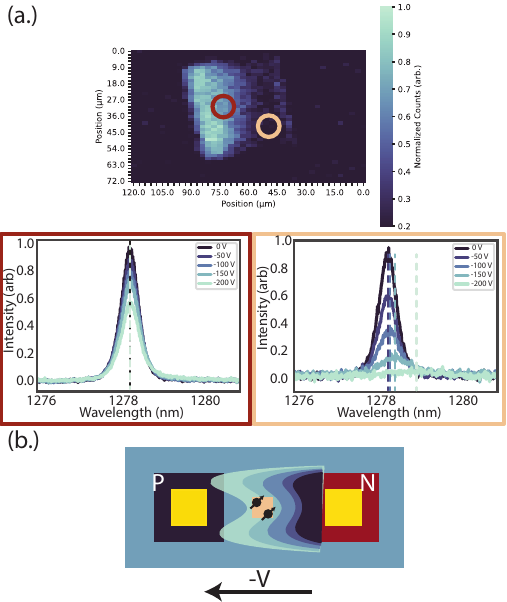}
\caption{\textbf{Variation in G center optical response across junction} \textbf{(a.)} The ensemble response is analyzed across the confocal PL map of Fig. \ref{Confocal}c from 0 and to -200~V. Close to the n-contact (brown), the ensemble experiences complete optical modulation and wavelength tuning--with tuning shown as distance between dotted lines, where the full data set is given in Fig. 3. In the middle of the junction (red), the ensemble experiences partial optical modulation with no wavelength tuning. \textbf{(b.)} Pictorial illustration of observed phenomena as larger reverse-bias voltage is applied. The local electric field accumulation and junction depletion reaches the ensemble (center orange square) at sufficient threshold voltage, resulting in wavelength tuning.}
\label{extinction}
\end{figure}

Furthermore, although partial optical modulation was observed at the center of the junction, wavelength tuning was not (Fig. \ref{extinction}a red). This finding is consistent with those reported experimentally in \cite{anderson2019}, and theoretically in \cite{Candido2021}, where at reverse bias voltages below a critical value, the size of the depletion region is less than the width of the junction. In \cite{anderson2019}, the threshold voltage to observe Stark effect of single divacancies in 4H-silicon carbide positioned at different spatial planes of a vertical diode depended on the position of the emitter in the junction. Here, we extend this argument by directly imaging the spatial dependence of the entire diode depletion region. Above a spatially-dependent threshold voltage where the junction depletion reaches the ensemble, a continual redshift of approximately 1.4~GHz/V is observed (Fig. \ref{extinction}a brown) \cite{SI}. However tens of microns away, where the junction depletion has not yet reached the ensemble, no wavelength tuning is experienced (Fig. \ref{extinction}a red). From these findings, precise determination of the G center differential polarizability is obfuscated due to the non-uniform field distribution in the junction and the distribution of dipole orientations within the ensemble \cite{redjem2023}. However future work using single emitters in an undoped \textit{I} layer of a PIN diode would enable this estimation to compare with theoretical predictions of the permanent dipole moment \cite{Ivanov2022}, as the precise Stark shift rate would be clearly captured by a single emitter and the lack of residual dopants would result in improved electric field uniformity. Wavelength-tuning is only observed in regions that also exhibit strong photocurrent, indicating the presence of the junction depletion region and large local electric field. This observation is conceptually illustrated in (Fig. \ref{extinction}b). These results suggest the Stark effect could be responsible for the observed emitter red-shift. The boron dopants in these areas within the junction are sufficiently depleted such that electric field can build up to yield Stark-shifted G centers. 

Finally, G center optical intensity is modulated both within and outside of the depletion region under increasing reverse bias. This observation could be explained by considering the effects of band-bending. The G center is thought to possess a bi-stability in its atomic configuration between an optically-active B configuration and a dark A configuration \cite{Song1990}. Optical emission arises in the transition from the meta-stable A to the B form under photo-injection \cite{Song1990,Beaufils2018,Udvarhelyi2021}. G center ensemble brightness thus depends upon the concentration of B configuration emitters: for lightly p-doped silicon at 4~K, the concentration of A configuration emitters should dominate B configuration emitters. However, optical excitation enables conversion of A configuration emitters to the metastable B configuration, where they can photoluminesce. Further, both A and B configuration emitters can be ionized to non-emissive charge states as the fermi level is tuned under external bias \cite{Song1990, Udvarhelyi2021}. Our observations are consistent with this explanation: as we increase the reverse bias across the junction, emitters in the A configuration convert to the emissive B configuration, but are probabilistically ionized to a dark state as a function of the resultant band bending.

\section{Conclusion}
We probed the coupling of a telecommunication-band silicon color center to DC electric fields by integrating G centers into diodes while retaining optical access. We then utilized the electrical manipulation of the ensemble to image the electric-field distribution within the diode, capturing the spatial evolution of the junction depletion region across varied reverse-bias voltages. Within the junction depletion region the ZPL redshifted by approximately 100~GHz at a rate of 1.4~GHz/V above a threshold voltage, whereas only modulation of the ZPL fluorescence intensity is observed outside of the depletion region. These findings suggest distinct emitter-field couplings are exhibited--with a spatial dependence across the junction--where band-bending and Stark effect could explain the observed phenomena. Furthermore, we find that hydrogen plays a critical role in the ability to observe G centers in our devices. To this end, future work will continue to elucidate the specific mechanisms involved in G center formation and stabilization, both via hydrogenation and electrical control. 

These devices provide a tool for electrically manipulating color centers with broad applicability to both other silicon color centers, and color centers in other semiconductor platforms. These findings using an ensemble of color centers to illustrate the spatial distribution of emitter-field coupling in the junction will motivate and inform the design of electrical devices to optimally couple to a single emitter. It would be of particular interest to observe the response of silicon T centers to electrical tuning via diode, as T centers possess a coherent spin-photon interface \cite{higginbottom2022}, and are reported to follow a similar synthesis procedure as was implemented here. Furthermore, our demonstration of the direct visualization of electric field dynamics in a semiconductor--optically mapping a DC electric field in-situ--has application in quantum sensing of electric fields \cite{dolde2011}. Finally, our lateral diode design at a buried plane of 110~nm is compatible with photonic crystal cavity integration \cite{bracher2017,crook2020}, where future work intends to enable simultaneous electrical tuning, stabilization, and control of cavity-enhanced quantum emitters.

\section{Methods}
\subsection{Fabrication}
All carbon, hydrogen, boron, and phosphorus ion implantation was performed at INNOViON Corporation. Ion implantation energies are determined using Stopping Range of Ions in Matter (SRIM) calculations \cite{SI}, targeting a depth of approximately $110~\text{nm}$ for each ion. Dopant densities are selected to obtain an acceptor/donor concentration of 1$\times 10^{19}$/cm$^3$ at the desired depth, as this order magnitude is typical of electrical devices in silicon. Further, overlapping the maximum dopant concentration depth with the etch-defined metalization ensures transmissive metal-semiconductor interface for ohmic contact. Each implantation was performed at a $7^{\circ}$ tilt. All masked implantation utilized optical lithography in the positive photoresist mask S1813 at a fluence of 250~mJ and wavelength of 375~nm using a Heidelburg Maskless Aligner 150. The resist was pre-baked at $115^{\circ}C$ for 3 minutes, and developed for 70 seconds in TMAH-based CD-26. Every photoresist mask was stripped with a 500~W $O_2$ plasma, and the Ti-Au resist-on-liftoff mask was stripped with a 12~hr soak in remover PG at $80^{\circ}C$.

First, an unmasked bulk wafer fragment is implanted with 7$\times 10^{13}$/cm$^2$ $^{12}$C ions at an energy of 38~keV. Next, $500\times500~\mu$m apertures are written in a photoresist mask with optical lithography, and 1$\times~10^{14}$/cm$^2$ $^{11}$B ions are implanted at an energy of 29~keV through the apertures to define localized p-doped islands. After resist stripping, n-doped islands are generated by implanting 1$\times~10^{14}$/cm$^2$ $^{31}$P ions at an energy of 80~keV through offset 500$\times$500$~\mu$m apertures again defined with optical lithography. The spacing between the p- and n-doped apertures (Fig. \ref{Fab}b) is swept across the wafer to vary the strength of the junction electric field for a given voltage. To both heal the crystal lattice and incorporate the dopants substitutionally in the silicon lattice \cite{Canham1987}, a rapid thermal anneal (RTA) is performed at $1000^{\circ}$C for 20 seconds in an argon environment after stripping the resist. 

Next, electrical contacts are generated by first writing 250$\times$250$~\mu$m apertures in a new resist mask positioned such that each opening was aligned to the center of the implanted dopant islands. Using SF$_6$ and C$_4$F$_8$ chemistry in a reactive ion etching chamber, the exposed windows are then etched down 220~nm to the oxide to ensure optimal overlap of the metal contacts with the implanted dopants. Following definition of a new 300$\times$300$~\mu$m aperture mask of photoresist on lift-off (S1813 on LOR3A), also aligned to the center of the implanted dopant islands, a thin film of 300~nm gold on a 30~nm titanium adhesion layer (Ti-Au) is deposited via electron beam evaporation. 

To complete the incorporation of G center ensembles, hydrogen is implanted through a window at the center of each junction (Fig. \ref{Fab}b). 7$\times~10^{13}$/cm$^2$ H ions were implanted at an energy of 9~keV through 50$\times$50$~\mu$m apertures in a final resist mask, forming an ensemble of diode-integrated G centers. The wafer fragment was subsequently diced into 6$\times$6~mm samples that were integrated into a 16-pin electrically-wired cryogenic cold-finger and wire-bonded for external driving (Fig. \ref{Intro}a).

\subsection{Experimental Setup}
Experiments are performed in a home-built confocal microscope using a Mitutoyo 100$\times$ 0.5~NA Near-IR objective. G centers are optically excited using an off-resonant $532~\text{nm}$ diode-pumped solid-state laser, and junctions are biased using a $\pm210~V$ Keithley 2400 source meter. Simultaneous optical and electrical measurements are enabled in a Janis ST-500 continuous-flow Helium-cooled cryostat with a 16-pin mapped electrical feed-through wire-bonded to the diodes. The system achieves a base temperature of roughly 6~K. Photoluminescence of the diode-integrated color centers is measured on an Acton Spectra Pro 2750 spectrograph with a Princeton Instruments OMA:V indium-gallium-arsenide nitrogen-cooled photodiode array detector. Raman spectroscopy is performed in a LabRAM Evolution Horiba multi-line room-temperature confocal Raman spectrometer using 532~nm laser excitation.

\section{Author contributions}
\noindent \textit{Methodology}: A.M.D., M.S., J.R.D., D.D.S., M.K.B., E.L.H; \textit{Fabrication}: A.M.D., A.R.; \textit{Measurement}: A.M.D., J.R.D.; \textit{Analysis}: A.M.D., M.S., J.R.D., D.D.S., M.K.B., E.L.H; \textit{Advising}: D.D.S., M.K.B., E.L.H; \textit{Manuscript Preparation}: All authors.

\section{Acknowledgments}
\noindent This work was supported by AWS Center for Quantum Networking and the Harvard Quantum Initiative. Portions of this work were performed at the Harvard University Center for Nanoscale Systems (CNS); a member of the National Nanotechnology Coordinated Infrastructure Network (NNCI), which is supported by the National Science Foundation under NSF award no. ECCS-2025158. M.S. acknowledges funding from a NASA Space Technology Graduate Research Fellowship.

\section{Data Availability}
\noindent The data that support the findings of the work are available from the corresponding author upon reasonable request.

\bibliography{SiPN.bib}
\end{document}

\title{Supplementary Information for ``Electrical Manipulation of Telecom Color Centers in Silicon"}
\date{\today}
\maketitle
\renewcommand{\thefigure}{S\arabic{figure}}

\tableofcontents
\clearpage

\section{Reverse Bias Extended Data}
\subsection{Depletion Region}
\begin{figure*}[h!]
\includegraphics[scale = 1]{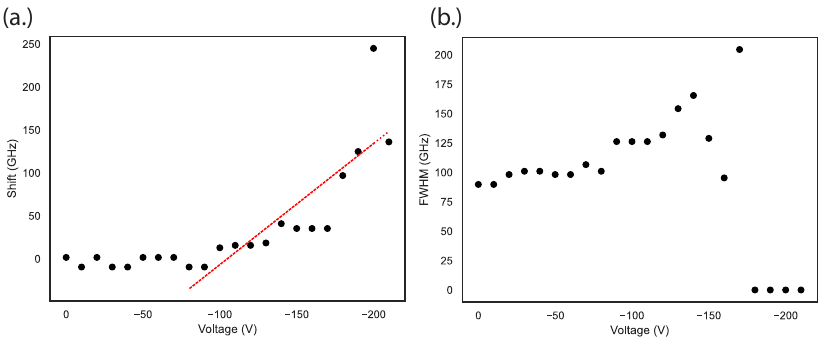}
\caption{\textbf{Reverse-bias within depletion region} Evolution of G center ZPL \textbf{(a.)} center wavelength redshift, and \textbf{(b.)} ensemble linewidth broadening from 0 to -210~V}
\label{Fig1}
\end{figure*}

Here we present photoluminescence center wavelength shift and ensemble linewidth from the main text Fig. 3 plotted as a function of applied voltage. At approximately 90~V reverse bias, the built in electric field reaches the interrogated region of the ensemble, and center wavelength redshift becomes evident. A linear fit is shown in red in  Fig. \ref{Fig1}a, providing an approximation of the tuning rate of 1.4~GHz/V, above the observed threshold voltage. The ensemble redshifts up to approximately 100~GHz until 100\% extinction of the fluorescence intensity is observed, as was shown in the main text. Similarly, the ensemble experiences broadening in response to applied voltage (Fig. \ref{Fig1}b). 

One potential explanation for the seeming discontinuities in the redshift and broadening trends could be the distribution of dipole orientations of individual emitters which comprise the ensemble tuning relative to each other in response to the electric field. In \cite{redjem2023}, it was found that an investigated G center ensemble possessed a distribution of dipole orientations with occurrence clustering around 20$\degree$ and 100$\degree$ detuned from the $\hkl<110>$ crystallographic axis. As mentioned in the main text, repeating our experiments with single emitters would provide clarity on our present observations. 

\clearpage
\subsection{Outside Depletion Region}
\begin{figure*}[h!]
\includegraphics[scale = 1]{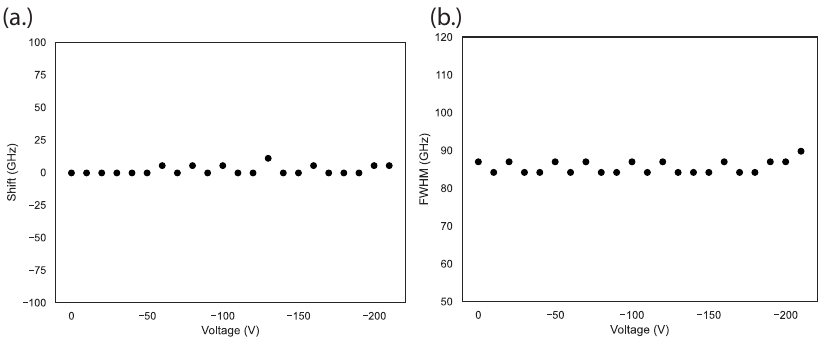}
\caption{\textbf{Reverse-Bias outside of depletion region} Evolution of G center ZPL \textbf{(a.)} center wavelength redshift, and \textbf{(b.)} linewidth broadening from 0 to -210~V}
\label{Fig2}
\end{figure*}

Similarly, the photoluminescence redshift and ensemble linewidth of the partial extinction region shown in red in main text Fig. 5a is plotted in Fig. \ref{Fig2} as a function of applied voltage. The fluorescence intensity of the ensemble reduces while experiencing no meaningful wavelength tuning or broadening between zero and maximum reverse bias. This can be explained by the absence of a local electric field interacting with the ensemble to yield a Stark shift, as the depletion region has not yet reached the investigated spatial region far from the n-contact. However, emitter ionization is still facilitated under band-bending, resulting in the observed reduction in ensemble brightness. 

\clearpage
\section{Forward Bias}
\begin{figure*}[h!]
\includegraphics[scale = 1]{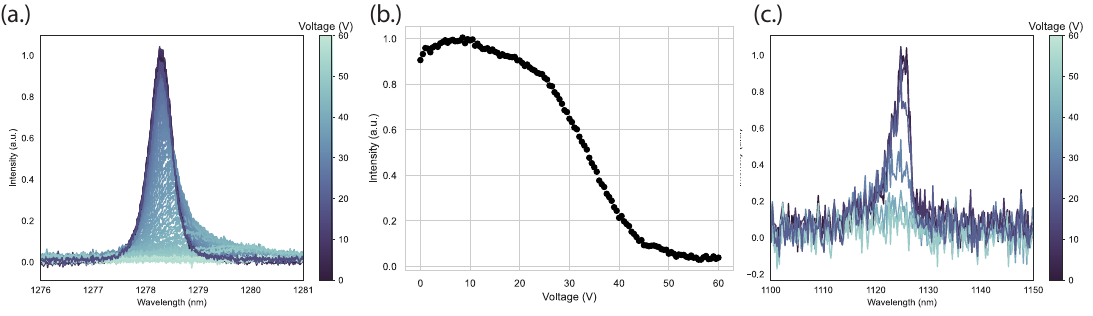}
\caption{\textbf{Forward-Bias response of G center ZPL and Silicon Free-Exciton} Evolution of \textbf{(a.)} G center ZPL with  \textbf{(b.)} modulation per voltage. \textbf{(c.)} Silicon Free-Exciton under application of a forward-biased electric field from 0 to 60 and 50~V, respectively. As expected, both the ensemble and exciton experience commensurate suppression at high current passage, unlike in the reverse-biased regime.}
\label{Fig3}
\end{figure*}

In contrast to the emitter-field coupling illustrated under reverse bias (Fig. 3 main text), in Fig. \ref{Fig3} we show that the optical response under forward bias--where orders magnitude greater current is passed-- is dominated by local heating within the junction. The optical response of the same diode-integrated G center ensemble and silicon free-exciton is analyzed under application of a 0 to 60~V forward bias (limited to avoid device degradation at high current passage). Notably, G center ZPL and exciton experience \textit{commensurate} modulation of the fluorescence intensity at increased forward bias--as to be expected at elevated local temperature (see SI Section 3). This finding is distinct from that of the reverse-bias results of the main text, where the silicon free-exciton is unperturbed while the ZPL experiences 100\% modulation. 

\begin{figure*}[ht!]
\includegraphics[scale = 1]{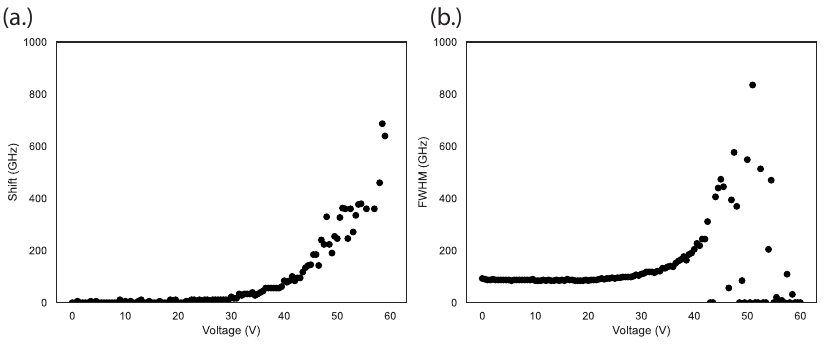}
\caption{\textbf{Wavelength Analysis of G center ZPL under Forward-Bias} \textbf{(a.)} center wavelength red-shift, and \textbf{(b.)} linewidth broadening, revealing order magnitude greater broadening and redshift occurring at lower absolute voltage than in the reverse-bias regime.}
\label{Fig4}
\end{figure*}

Similarly, in Fig. \ref{Fig4}, the G center redshift and linewidth evolution undergo markedly different qualitative responses when compared to the reverse bias case. Here ten-times greater linewidth broadening and wavelength-tuning is experienced at 4-times lesser absolute voltage. Under large forward bias power, the emitter-field coupling competes with local heating in the junction, resulting in ZPL broadening and shifting consistent with that observed under direct heating. These results capture the optimal electrical performance of the junction, and further illustrate the unique behavior observed above under reverse-bias in the main text. 

\clearpage
\section{Direct Stage Heating}
\begin{figure*}[ht!]
\includegraphics[scale = 0.65]{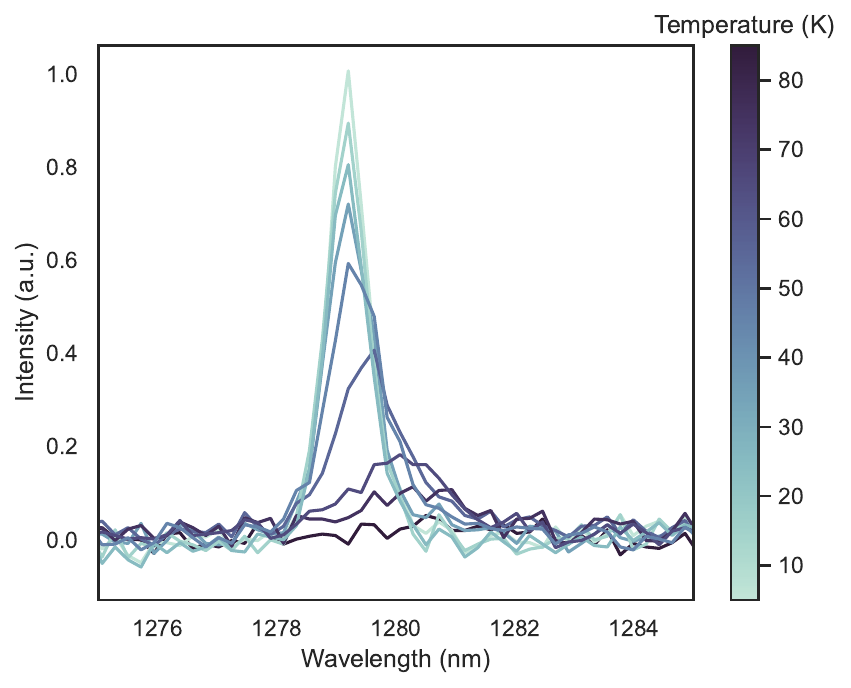}
\caption{\textbf{G center ZPL evolution under direct sample stage warming} \textbf{(a.)} G center ZPL evolution from roughly 4~K to 80~K. As a function of temperature, the ensemble optical efficiency, linewidth, and center wavelength undergo a simultaneous reduction, broadening, and redshift, respectively. At roughly 80~K, one second integration on the spectrometer is insufficient to resolve the emitter ZPL.}
\label{Fig5}
\end{figure*}

For reference, heating power is directly applied to the sample stage cold finger in the cryostat on a bulk reference sample (no electrical driving of diode-integrated emitters), and the G center ensemble photolumninescence is measured as a function of cryostat temperature (In Fig. \ref{Fig5}). The ZPL evolution follows notably similar reduction, broadening, and redshift as is reported above via forward biasing an electrical junction. Comparing the ZPL intensity and linewidth under direct sample stage heating to that of the forward biasing of electrical junctions, it appears that the ensemble is elevated to approximately 60~K. At 50~V forward bias, approximately 9~mA is passed through the junction--corresponding to 450~mW power. Therefore, we can estimate that forward-bias driving the junction with roughly 27~dBm locally heats the ensemble by 55~K. 

\clearpage
\section{Hydrogenation}
Here we investigate optimal means of G center formation and device integration. The optical and electrical properties of the devices and the formed emitters are studied under varied methods of hydrogen inclusion. Specifically, we vary the hydrogen ion implantation densities, and then compare with samples exposed to a hydrogen plasma of incremental duration ranging from zero to two minutes. The plasma is generated in a reactive ion etching chamber solely comprised of H$_2$ gas under 35~mTorr pressure, a flow-rate of 30~sccms, and using 1000~W coil and 200~W platen powers.

\subsection{SOI Raman Spectroscopy}
\begin{figure*}[ht!]
\includegraphics[scale = 1]{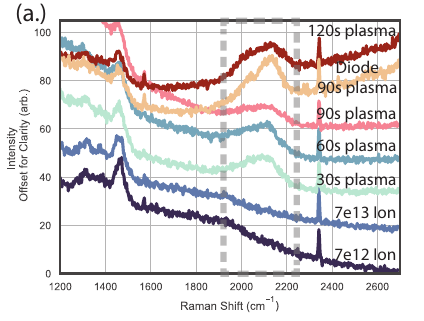}
\caption{\textbf{SOI Raman} \textbf{(a.)} Raman spectra comparing hydrogen ion implantation to hydrogen plasma exposure across SOI samples. The $2100cm^{-1}$ Si-H stretch bond (dotted-gray boxed region) is only present for samples which received hydrogen plasma exposure.}
\label{Fig7a}
\end{figure*}

To understand the extent of hydrogen incorporation during implantation versus plasma treatment, the Raman signatures of samples that underwent each treatment are shown in Fig. \ref{Fig7a}. All samples were first carbon ion implanted and rapid thermal annealed at 1000~C for 20~s. One sample additionally possessed the diode devices discussed throughout this work. The Raman spectra of all investigated samples have the same basic features with the exception of either the presence or absence of a peak at $2100cm^{-1}$. This peak has previously been identified as characteristic of a silicon-hydrogen (Si-H) stretch-bond present on sample surfaces \cite{brodsky1977}. This Si-H stretch bond line is evident in all samples which were exposed to the hydrogen plasma, even those which received the shortest investigated exposure time of 30~seconds. However, this peak was absent from the samples which received hydrogen implantation and those which were untreated, suggesting the plasma treatment uniquely modifies the wafer surface. 

\subsection{Float-Zone H Plasma Spectra}
\begin{figure*}[ht!]
\includegraphics[scale = 0.5]{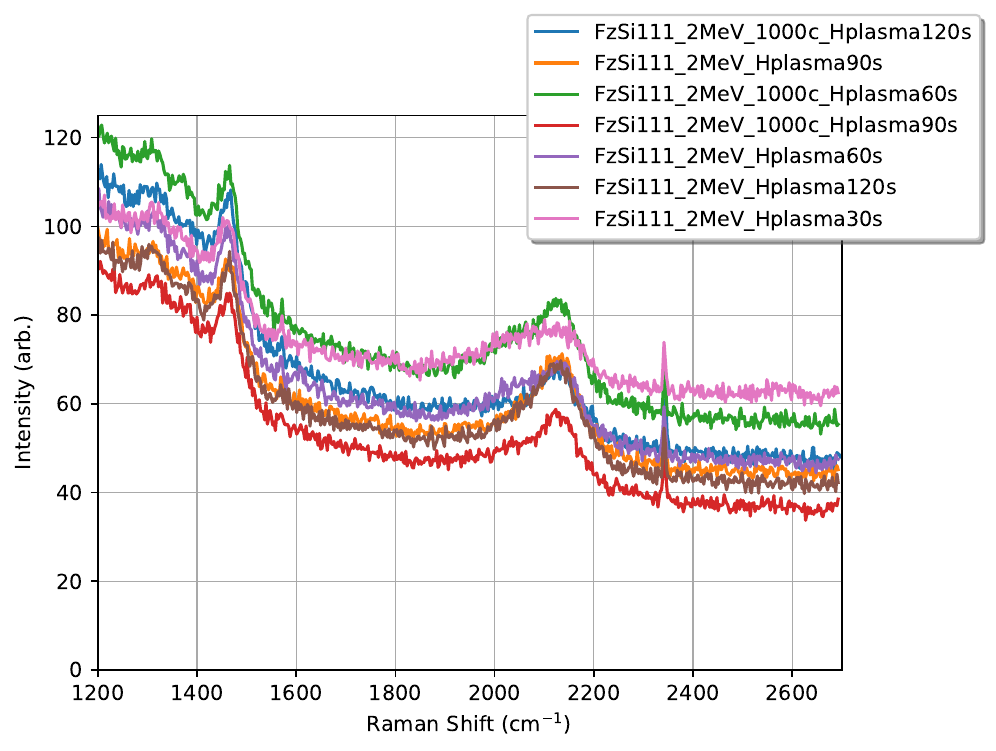}
\caption{\textbf{Float Zone Si Raman} \textbf{(a.)} Evidence of Si-H surface raman bond in float zone silicon analyzed under varied plasma exposure}
\label{Fig7}
\end{figure*}

The effect of hydrogen implantation versus plasma treatment is also investigated on electron-irradiated float-zone (FZ) silicon samples. Similarly, the $2100cm^{-1}$ peak occurs only in samples that underwent hydrogen plasma exposure (Fig. \ref{Fig7}).

\begin{figure*}[ht!]
\includegraphics[scale = 0.55]{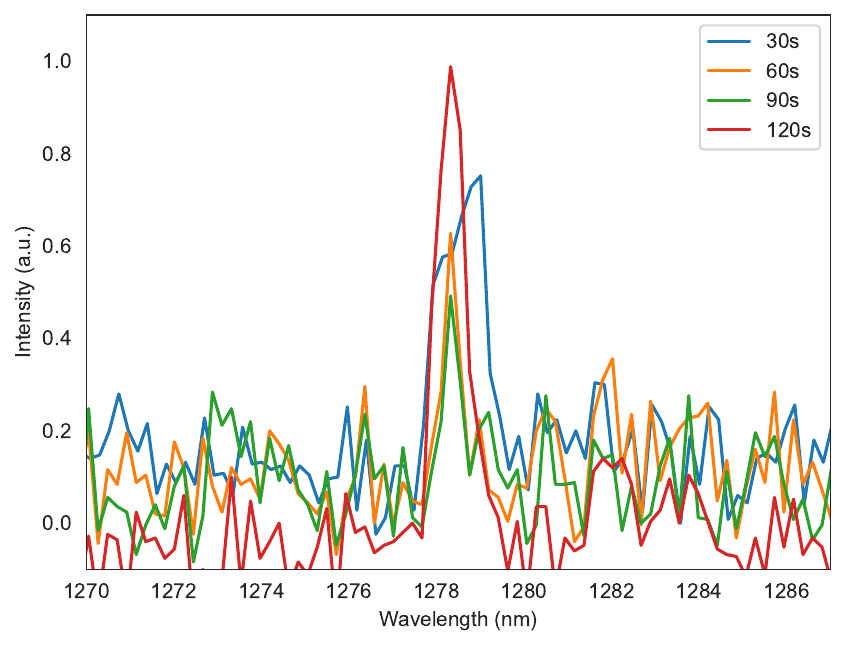}
\caption{\textbf{Float Zone Si PL} \textbf{(a.)} Associated G center PL of FZ samples exhibiting the Si-H raman bond.}
\label{Fig8}
\end{figure*}

Additionally, In \ref{Fig8} we report the correlated observation of G center PL in FZ samples which received the above hydrogen plasma treatment. These findings indicate that surface-related hydrogenation is sufficient to enable optical observation of G centers. Previous reports have considered bulk proton irradiation, however here we show that even a 30~second exposure of a surface treatment renders observation of G centers. 

We thus establish both the correlation of the Si-H Raman band with hydrogen plasma exposure, and the ability to form G centers with both bulk and surface hydrogenation. Taken together, these results suggest that hydrogenation may play a central role in forming or stabilizing G center emission. However, further investigation is required to understand the contributory atomic-scale dynamics and mechanisms. For instance, though the plasma treatment is shown to be a surface-related effect by the Raman signature, it is not yet clear whether physical or chemical mechanisms render the formation of G centers under hydrogen plasma exposure--as plasma-induced knock-out could still mediate the incorporation of interstitial carbon for shallow G center formation on the FZ surface. 

\subsection{Localized hydrogen incorporation}
Though the presence of hydrogen was found to be critical to generate G centers, the interaction of hydrogen with the fabricated devices must also be considered. Significant electrical degradation is observed in diodes which received either blanket hydrogen ion implantation or blanket plasma exposure (see Supplementary Information Section 5). Junction resistance increased as a function of hydrogen plasma exposure, as shown in the IV-curves of Fig. \ref{Fig9}. Further, the blanket-implanted devices no longer passed appreciable current at cryogenic temperatures Fig. \ref{Fig10}. The utilized p-dopant, boron, is known to be passivated by hydrogen, and similarly, the conductivity of the gold contacts is expected to degrade under hydrogenation. To avoid these issues, a mask defined by optical lithography is ultimately employed for the utilized paper devices to achieve the formation of the desired emitters within the fabricated devices while preserving the electrical behavior of the diodes. 

\clearpage
\section{Electrical Analysis under Blanket Hydrogenation}
In the main text, it was reported that masked hydrogenation was required to render simultaneous observation of G centers and the preservation of the electrical performance of the diodes. Here we report the data observed under various studies which motivated this decision. 
\subsection{Hydrogen Plasma}
\begin{figure*}[ht!]
\includegraphics[scale = 1]{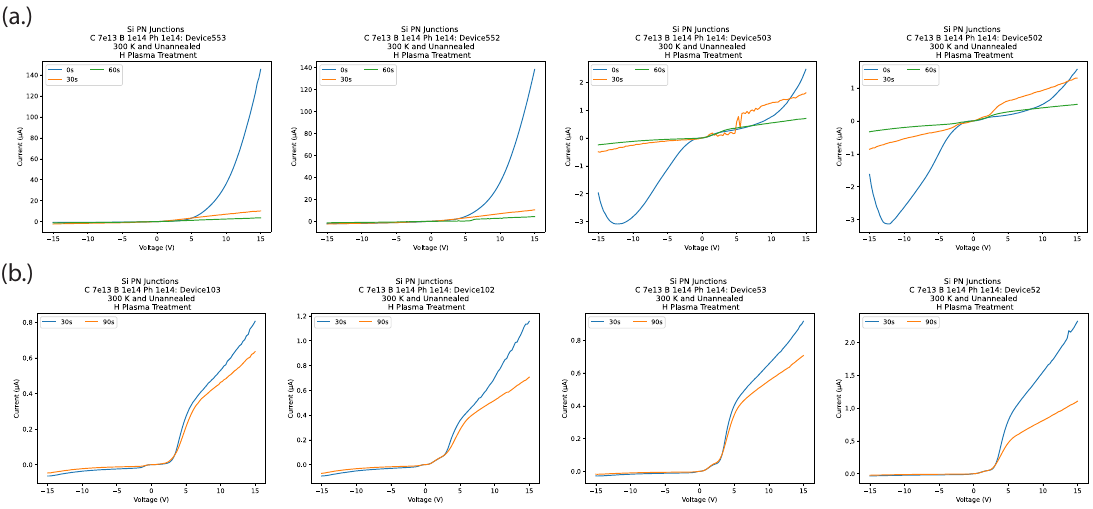}
\caption{\textbf{IV-analysis under H plasma} Consistent degradation of diode characteristics under increased hydrogen plasma exposure, evidence across every investigated device. Evolution of eight different devices are given, with \textbf{(a.)} successive 30 second treatments, and \textbf{(b.)} a 30 second treatment followed by a 60 second treatment.}
\label{Fig9}
\end{figure*}

The electrical properties of diodes are investigated under incremental exposure to a hydrogen plasma. A continual reduction of current passage--and seemingly increase in contact resistance--is observed for each subsequent exposure in every investigated device. Fig. \ref{Fig9}a shows room temperature IV-curves of four devices under 30 second increments of the hydrogen plasma exposure discussed in the main text, ranging from 0-60 seconds. In Fig. \ref{Fig9}b, we repeat this process on four new devices, this time allowing a full minute of exposure, such that these devices received first a 30~s and then a 60~s exposure. The peak current passage is uniformly reduced for all investigated conditions on all evaluated devices, consistent with an increase in contact resistance, and potential passivisation of boron dopants in the junction. 

\clearpage
\subsection{Ion Implantation}
\begin{figure*}[ht!]
\includegraphics[scale = 0.5]{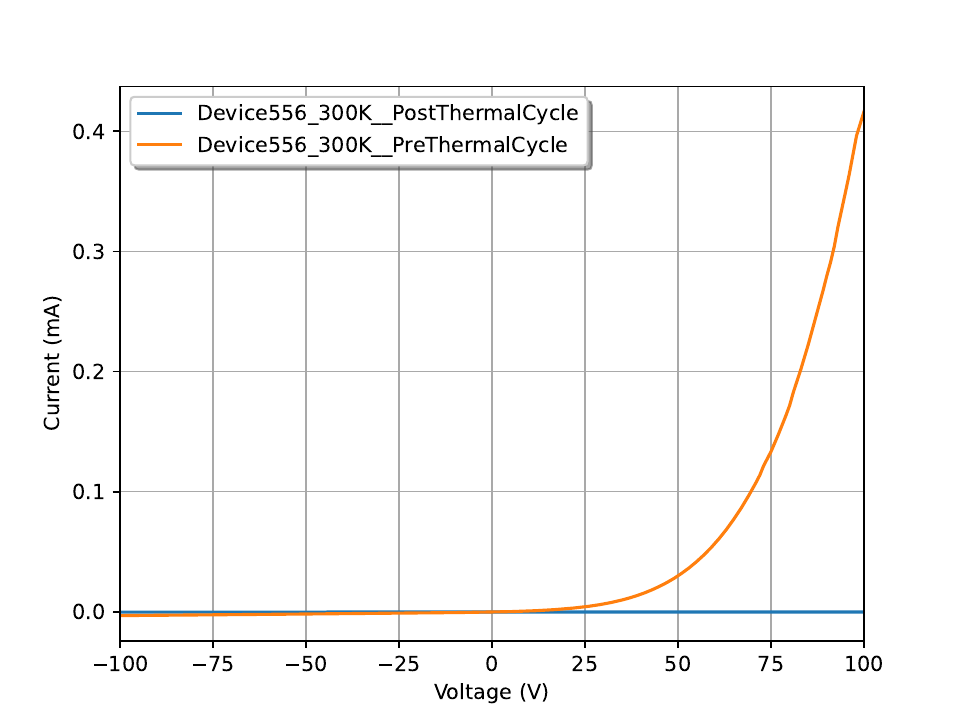}
\caption{\textbf{Ion implanted IV-analysis under cooling cycle} \textbf{(a.)} Diodes which received a blanket hydrogen ion implantation suffered permanent damage upon a thermal cycle in the cryostat. Furthermore, current passage is heavily reduced (0.4~mA at 100~V) compared to the device of the main text which received a masked implant.}
\label{Fig10}
\end{figure*}

Diodes which were blanket implanted with hydrogen ions showed a similar degradation of their electrical properties following implantation as is evident under incremental plasma treatment. However, upon a 4~K cooldown cycle, the devices suffered permanent damage where appreciable current is no longer passed at all (Fig. \ref{Fig10}). The pre-cooldown operation of the device is not recovered upon warm-up, and thus it appears that the process of cooling the bulk hydrogenated diodes imparts permanent damage--perhaps due to contact delamination under thermal expansion and contraction upon hydrogenating the Ti-Au.

\clearpage
\section{G center annealing stability}
\begin{figure*}[h!]
\includegraphics[scale = 0.5]{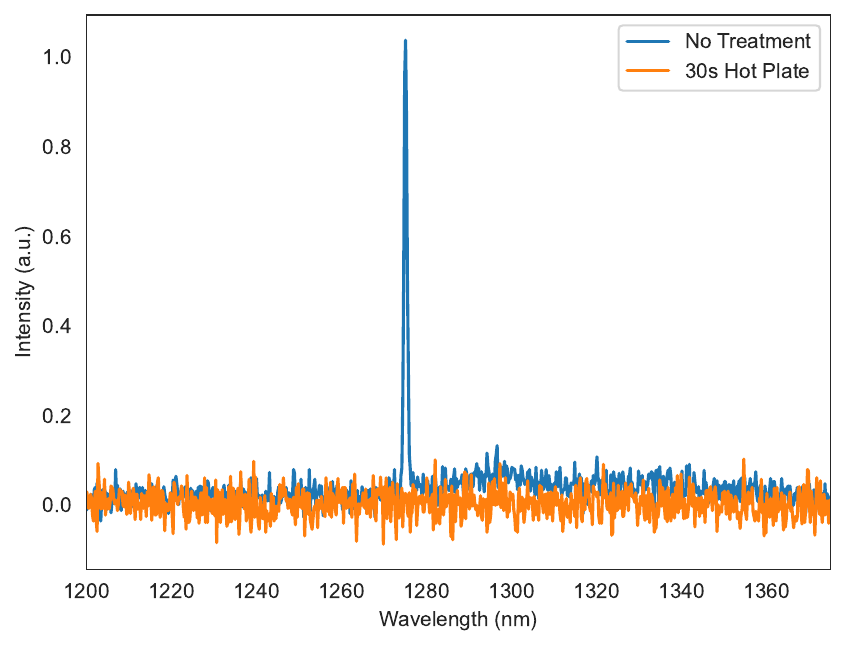}
\caption{\textbf{Time-dependent analysis of rapid thermal anneal} \textbf{(a.)} All investigated times of exposure to a 400$\degree$C hot plate resulted in quenched G center emission, even in as little as 30 seconds. A $4\times4$~mm wafer sample was directly placed on a thermally-equilibriated hot plate for exactly 30 seconds, then removed and allowed to naturally cool.}
\label{Fig6}
\end{figure*}

We analyze the stability of optical emission of G center ensemble under varied annealing conditions to best evaluate emitter integration in the fabricated electrical devices. Here the investigated samples are all identically co-processed up until the thermal anneal, consistent with the emitter synthesis outlined in the main text. Then, each sample is placed on a hot plate at 400$\degree$C for incremental times. Notably, we observe complete loss of optical emission of the ensemble in as short at 30 seconds of heating (Fig. \ref{Fig6}). 

Additionally, we repeat the experiment on samples held at lower temperatures and find that G center emission is unaffected. We observe preserved optical emission in samples held on a hot plate at 115$\degree$C for 3min, and at 180$\degree$C for 3min. These two temperatures and times are investigated due to their association with with resist mask baking for optical and electron-beam lithography, respectively. 

In summary we find G center optical emission to be stable on samples annealed up to 180$\degree$C, but unstable at 400$\degree$C under as little as a 30~s anneal. These findings are consistent with those reported in \cite{Canham1987}, where they find G center optical emission to be unstable above 200$\degree$C. For this reason, incorporation of G centers using hydrogen is performed as the final step in the fabrication process. Similarly, future work using samples which received a contact-anneal will incorporate hydrogen following the contact anneal, as gold contact annealing temperatures (investigated below in SI Section Seven) are around 400$\degree$C. 

\clearpage
\section{Electrical Contact Annealing}
\begin{figure*}[ht!]
\includegraphics[scale = 0.65]{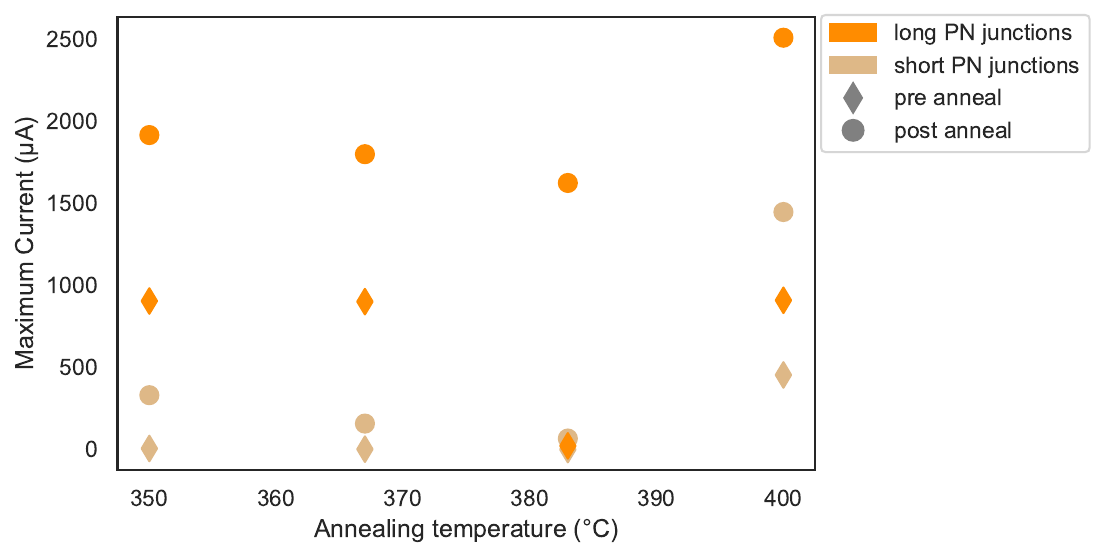}
\caption{\textbf{Summary of contact anneal enhancement} \textbf{(a.)} Diodes which received a contact anneal exhibited uniform improvement in their IV-characteristics at all investigated temperatures from 350-400$\degree$C. All investigated devices show improvement from anneal, and the degree of enhancement is approximately $2\times$ for most of the investigated devices.}
\label{Fig11a}
\end{figure*}

\begin{figure*}[ht!]
\includegraphics[scale = 0.5]{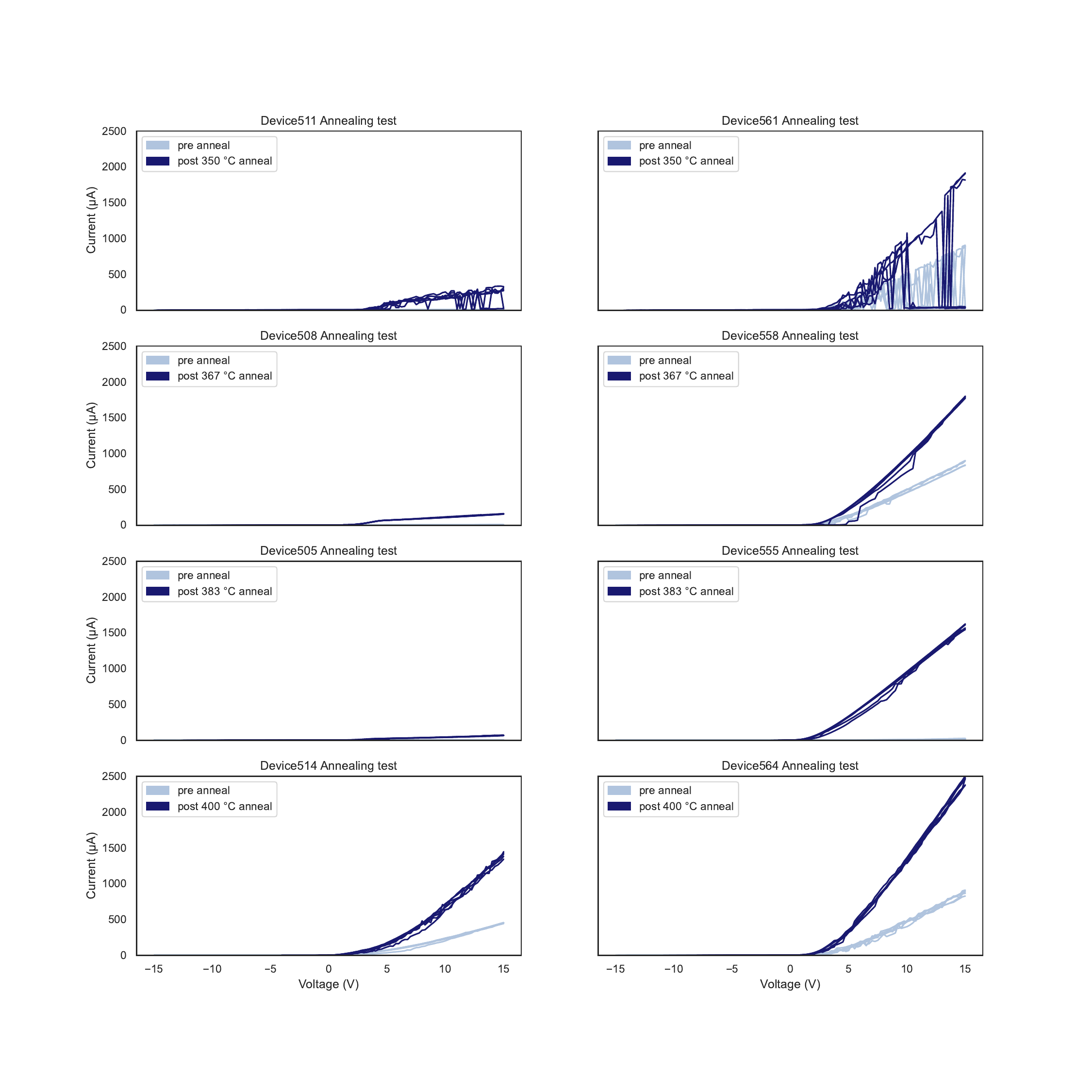}
\caption{\textbf{Detail of contact anneal} \textbf{(a.)} IV curves are measured before and after a contact anneal at four different temperatures for eight different devices. The left column are \textit{short} junctions, where the right column are \textit{long} junctions. The data of these curves are summarized in Fig. \ref{Fig11a}}
\label{Fig11b}
\end{figure*}

The electrical performance of our fabricated diodes is investigated under thermal annealing at varied temperatures, with the findinds summarized in Fig. \ref{Fig11a}, taken from the data of Fig. \ref{Fig11b}. The thermal anneal is expected to improve the ohmic nature of the Ti-Au electrical contacts, and thus four different conditions are evaluated. Each anneal was performed in a tube furnace under vacuum, with the target temperature held for 2 minutes. The pressures of each anneal, from 350~$\degree$C to 400$\degree$C, were 7.1e-6, 8.2e-6, 9.5e-6, 6.6e-6 Torr, respectively. It took approximately 9 minutes for the furnace to ramp from 20$\degree$C to target temperature, and the sample cooled down naturally after the 2 minute anneal. We characterize the degree of improvement achieved via anneal by the maximum current passed through the junction for the given IV-sweep that was selected (-15 to 15~V), as enhanced electrical contact between the dopants and the electrode will result in lower contact resistance. 

We classify the response of our devices between \textit{short} and \textit{long} PN junctions, where \textit{short} refers to devices with a junction width between doped pads less than $50~\mu$m. The long junctions systematically out-perform the short junctions, including in the degree of enhancement found under thermal anneal. We suspect the reason for this is that for a small gap between the doped regions, mask alignment errors from the successive aligned optical lithography writes likely result in a partial overlap of the P and N dopant aperatures, resulting in poor diode performance. 

Ultimately, the device used in this work (a $103~\mu$m gap \textit{long} junction) received no contact anneal, as it's performance was sufficient to observe the desired emitter-field interaction. However, future works would benefit from performing a 2 minute thermal anneal at 400$\degree$C prior to hydrogen incorporation. 

Additionally, we note that the T center in silicon has been reported to be observed following a comparable annealing condition as was investigated here for the electrical contacts. As such, it would be interesting to investigate the diodes under the reported \cite{macquarrie2021} T center process of: \begin{enumerate}
    \item hydrogen ion implantation
    \item 1 hour boil in deionized water
    \item $\sim\ 400\degree$C rapid thermal anneal for simultaneous T center synthesis and contact annealing. 
\end{enumerate}
to find if both the electrical properties of the diodes improve and if T centers are observed. However, in our background investigations with bulk SOI, T centers are not observed following the reported recipe. 

\clearpage
\section{COMSOL Simulations}
\begin{figure*}[h!]
\includegraphics[scale = 1]{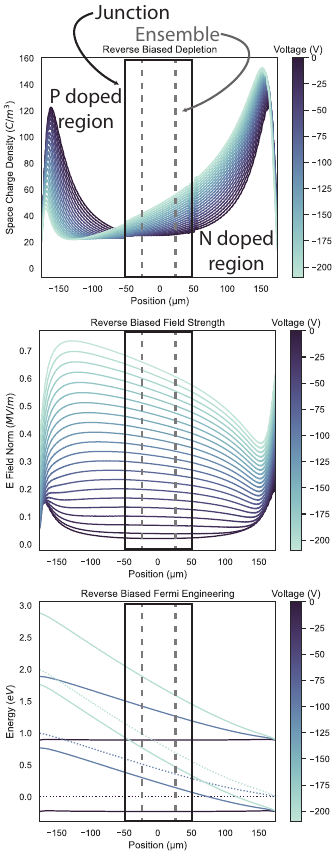}
\caption{\textbf{COMSOL simulation of Paper Device} Analysis in spatial region occupied by the G center ensemble, at a depth of 110nm, of \textbf{(a.)} space charge density, \textbf{(b.)} Electric field norm, and \textbf{(c.)} valence and conduction band edge bending (with electron quasi fermi level shown as dotted line)}
\label{Fig12}
\end{figure*}

A COMSOL model of the device presented in the main text is built and simulated (Fig. \ref{Fig12}) using the semiconductor module to study the spatial dependence on the space charge, electric field strength, and band bending in the diode in the region where the G center ensemble is located. The investigated device has a device size \textit{d}--defined from center to center of the metal contacts--of $603~\mu$m. Therefore the gap between doped apertures that the junction and ensemble occupy is $103~\mu$m. 

Qualitative trends are evident from the simulations in which the junction depletion and electric field strength move toward the P contact with increased reverse bias voltage, and that the region of maximum space charge depletion coincides with strongest field. These findings are all confirmed by our experimental data in main text Fig. 5. Similarly, the band-bending is exhibited across the junction in the simulation, which could agree with the modulation of optical intensity observed experimentally. However, exact quantitative agreement with simulation is not strong, as the simulations less clearly capture the spatial evolution of junction depletion as is observed experimentally--where local electric field and junction depletion does not reach the ensemble until a sufficient threshold voltage. A more accurate simulation of these effects is calculated from first principles in \cite{Candido2021}--to validate similar experimental \cite{anderson2019} findings as ours--in which the relevant physics are more fully captured than with a simple COMSOL model. 

\clearpage
\section{SRIM simulations}
Stopping range of ions in matter (SRIM) simulations are performed for the four ion species used in this work to find the optimal energies and doses to incorporate boron, phosphorus, carbon, and hydrogen in the same spatial plane and at desired densities. 

\begin{figure*}[ht!]
\includegraphics[scale = 1]{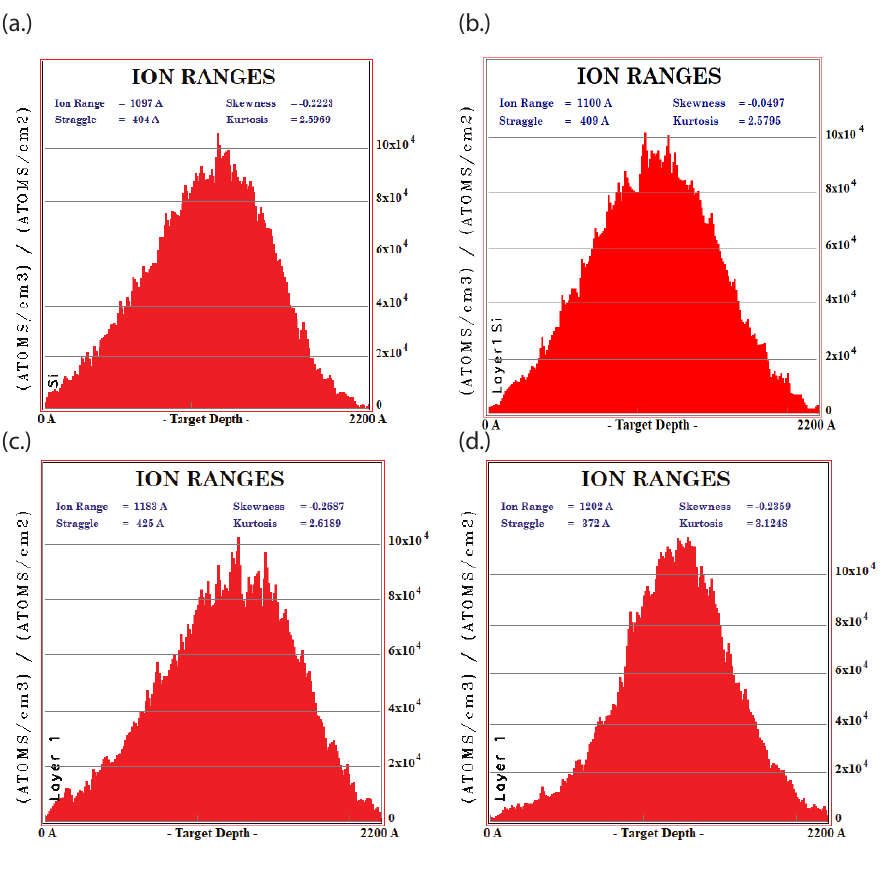}
\caption{\textbf{SRIM simulations of utilized ions} SRIM calculations to enable a stopping distance at 110~nm using \textbf{(a.)} boron, \textbf{(b.)} phosphorus, \textbf{(c.)} carbon, and \textbf{(d.)} hydrogen.}
\label{Fig13}
\end{figure*}

\bibliography{SiPN.bib}% Produces the bibliography via BibTeX.